**Title:** Ultrafast photocurrent measurement of the escape time of electrons and holes from carbon nanotube PN junction photodiodes

**Authors:** Nathaniel. M. Gabor[1,*], Zhaohui Zhong[2], Ken Bosnick[3], Paul L. McEuen[1,4]

[1]*Laboratory of Atomic and Solid State Physics, Cornell University, Ithaca, NY 14853, USA.*

[*]*Current Address: Department of Physics, Massachusetts Institute of Technology, Cambridge, MA 02139, USA.*

[2]*Department of Electrical Engineering and Computer Science, University of Michigan, Ann Arbor, MI 48109, USA.*

[3]*National Institute of Nanotechnology, National Research Council of Canada, Edmonton AB, T6G 2M9, Canada.*

[4]*Kavli Institute at Cornell for Nanoscale Science, Cornell University, Ithaca, NY 14853, USA.*

**Abstract:** Ultrafast photocurrent measurements are performed on individual carbon nanotube PN junction photodiodes. The photocurrent response to sub-picosecond pulses separated by a variable time delay $\Delta t$ shows strong photocurrent suppression when two pulses overlap ($\Delta t = 0$). The picosecond-scale decay time of photocurrent suppression scales inversely with the applied bias $V_{SD}$, and is twice as long for photon energy above the second subband $E_{22}$ as compared to lower energy. The observed photocurrent behavior is well described by an escape time model that accounts for carrier effective mass.

Highly efficient photovoltaic and photo-detector devices, which make use of multiple electron-hole pair generation from a single photon [1-4], require comprehensive understanding of charge carrier dynamics and their role in optoelectronic response. In order to study dynamics in nanoscale systems such as carbon nanotubes (NTs) and nanocrystal quantum dots, numerous measurements have been developed to probe the relevant time scales of electron motion in ensembles of these novel materials. In NTs, the time scale over which carrier interactions occur may range from $10^{-14}$ seconds for intrasubband relaxation [5] to greater than $10^{-9}$ to $10^{-7}$ seconds for radiative recombination [6-9]. However, no measurements have combined ultrafast optical and electronic techniques to probe the carrier dynamics and interactions in *individual* nanotube optoelectronic devices. While optical probes measure either the creation of electron hole pairs/excitons (absorption) or their relaxation (emission), photocurrent measurements probe a different quantity, the photoexcited carriers that escape the junction as separate electrons and holes. This time scale is not easily accessible from optical measurements, but is key for understanding the behavior of photovoltaics.

In this Letter, we present the first ultrafast photocurrent measurements of an individual NT optoelectronic device that incorporates sub-picosecond laser pulses. Using our technique, we directly probe the transit of electrons and holes through a NT PN junction in the time domain, finding that carriers in the first subband (of effective mass $m_1^*$) escape the device in half the time as carriers in the second subband ($m_2^* = 2m_1^*$). Our measurements indicate that carrier escape is diffusive in forward bias and, as the escape time decreases, approaches ballistic transport in reverse bias.

A schematic of the device is shown in Figure 1, as described previously [2,10,11]. Gate electrodes beneath a nanotube form a p-i-n junction, and the $I$-$V_{SD}$ curve of the device shows a diode characteristic (Figure 1(b)). The turn-on voltage gives an approximate measure of the bandgap $V_{OC} = E_{11}/e$, and standard photocurrent spectroscopy measurements (Figure 1(c)) can be used to measure the energy $E_{22}$ of the second subband [2,12,13]. For the device shown, referred to as device 1, these are found to be $E_{11} = 0.48$ eV and $E_{22} = 0.95$ eV.

Figure 1(a) shows the experimental schematic for measuring photocurrent at ultrafast time scales. A femtosecond Ti:Sapphire laser (<200fs pulse width and 75MHz repetition rate) or an Optical Parametric Oscillator, with respective wavelength ranges of 780-1000 nm and 1200-1600 nm, is used to photo-excite the nanotube PN junction. The beam is focused through a microscope objective onto the NT sample in an optical cryostat at low temperatures. We measure photocurrent response to single pulses or as a function of the time delay between two pulses. This is accomplished by splitting the output laser beam into a reference and delay beam separated by a time interval $\Delta t$. This temporal separation can be tuned by varying the optical path of the delay beam.

We first measure the low temperature photocurrent at the PN junction due to a single optical pulse train (at $f = 75$MHz) as a function of the excitation intensity $n$ (number of photons per pulse / cm$^2$). Figure 2 shows photocurrent vs. intensity at $V_{SD} = 0$ V for device 1. We normalize the photocurrent data by the current value at which one carrier is generated per pulse (inset Figure 2(a)): $I = ef \sim 12$ pA, where $e$ is the elementary charge and $f$ is the repetition rate of the laser. The photocurrent is linear for $I / ef < 1$ but becomes sublinear above $I / ef > 1$. The sublinear behavior can be approximately described as $I \sim n^{0.3}$ (Figure 2(b)).

The data of Figure 2 indicate that when multiple excitations dwell simultaneously in the junction, they strongly reduce the photocurrent response, likely due to electron-hole recombination. We can use this sublinearity of the photocurrent vs. intensity to probe the relevant time scale during which photo-excited excitations reside in the junction before escaping. In other words, how long must we wait before the junction is again empty? At zero time delay, two overlapping pulses will drive the photocurrent into strong sublinearity, while at long time delays the photocurrent will respond as though the pulses are independent, producing a larger current. The crossover between these two behaviors yields the escape time from the junction.

Figure 3 shows the double pulse photocurrent measured at $V_{SD}$ = 0 V and $E_{PH}$ = 1.51 eV for the same device as in Figure 2. In Figure 3(a), as intensity increases, we observe a photocurrent dip near $\Delta t$ = 0 ps (when the two pulses overlap). The photocurrent dip is symmetric at positive and negative time delay and has a temporal width of ~400 fs at low intensities (experimental detection limit) and saturates to ~1 ps at high intensities.

We normalize the high intensity photocurrent near $t$ = 0 (Figure 3(b)) and observe an exponential dependence vs. time delay with a characteristic decay time constant $\tau$ = 0.8 ps at $V_{SD}$ = 0 V. In the remaining sections, we discuss the dependence of the double pulse photocurrent decay time on source drain bias and photon energy.

Figure 3(b) compares the normalized photocurrent vs. time delay at $V_{SD}$ = 0 V and $V_{SD}$ = - 0.3 V. As the device goes from zero bias into reverse bias, the characteristic decay time $\tau$ decreases. We extract the characteristic decay constant at many $V_{SD}$ values and plot them in Figure 3(c). In reverse bias, the decay time remains constant $\tau_0$ ~ 0.5 ps (labeled with a solid blue line). As $V_{SD}$ approaches the open circuit voltage ($V_{OC}$ = 0.48 V), the decay constant $\tau$ increases rapidly to $\tau$ = 1.4 ps at $V_{SD}$ = 0.15 V. Due to the decrease of photocurrent as $V_{SD}$ approaches $V_{OC}$, characteristic time constants cannot be extracted close to $V_{OC}$. In the inset to Figure 3(c), we plot the inverse decay time $1/\tau$ as a function of $V_{SD}$. Importantly, the inverse decay time scales linearly with $V_{SD}$ with a negative slope of $|s|$ = 2.3 (V-ps)$^{-1}$ and extrapolates to an intercept of $V_{SD}$ = 0.45 V as $1/\tau$ approaches zero.

The $V_{SD}$ dependence of the decay time suggests that $\tau$ is set by the escape of electrons and holes out of the PN junction. After optical excitation, electrons and holes are separated in the built-in electric field $\mathcal{E}$ and accelerate towards the device contacts (Figure 3(d)). As the electric field increases (moving from the flat band condition at the open circuit voltage into reverse bias), the charge carriers escape more quickly.

One model to describe this behavior is diffusive transport. During their escape from the junction, electrons and holes generated at the center of the device must travel a distance $L$ with an electric-field dependent drift velocity $v_D = \mu\mathcal{E}$ where $\mu$ is the mobility. From the velocity, we get an expression for the escape time of electrons and holes out of the junction

$$\tau = \frac{2L^2}{\mu(V_{OC} - V_{SD})}.\tag{1}$$

Here, $\mathcal{E} = V/L$ is the electric field resulting from a voltage $V$ applied over a distance $L$. The total applied voltage is $V = (V_{OC} - V_{SD})/2 = (E_{GAP}/e - V_{SD})/2$, and $L$ is half the length of the device since electrons and holes are generated at the center.

Comparing Equation (1) to our data, we see that the linear fit in the inset of Figure 3(c) indeed extrapolates to the open circuit voltage $V_{OC} \sim 0.48$ V which gives the band gap energy $E_{11} \sim 0.48$ eV. We can measure the length of the junction region using scanning photocurrent microscopy [2] and find a total junction length of $\sim 1$ μm for this device.

Combining half the junction length $L \sim 0.5$ μm with the slope from Figure 3(c), we estimate the mobility in the *intrinsic* region of the PN junction: $\mu = s2L^2 = 2(2.3 \text{ (V-ps)}^{-1})(0.5 \text{ μm})^2 \sim 1$ μm$^2$/V-ps $\sim 10^4$ cm$^2$/V-s, which is comparable to mobility values measured in high-mobility NT devices [14,15]. We can also establish the upper limit of the scattering length of carriers as they transit the junction: $l \leq v_F \tau_S = v_F \mu m_2^*/e = \mu E_{22}/2ev_F \sim 0.5$ μm, where $\tau_S$ is the average time between scattering events and $m_2^* = E_{22}/2v_F^2$ is the second subband effective mass. This scattering length is comparable to the intrinsic region length $L$, indicating that the transport is at the border between diffusive and ballistic. It is slightly larger than the scattering length of high-energy ($\varepsilon_{OP} \sim 0.2$ eV) optic phonons [17-21], the emission of which occurs with mean free path $l_{OP} \sim 100$ nm in semiconducting nanotubes [21].

We can also compare the results to a ballistic carrier model in the PN junction. In NTs, carrier energies are given by a hyperbolic band structure in which the upper limit to the velocity of electrons and holes is the Fermi velocity $v_F \sim 0.8$ μm/ps [16,17,22,23]. In an electric field, ballistic transport is analogous to a relativistic electron in a static field limited by the speed of light. In the low-energy limit, the escape time varies inversely with $V_{SD}^{-1/2}$, analogous to a classical ballistic particle. This is not observed in Fig 3(c) and so rules out purely ballistic transport in forward bias.

One prediction of the diffusive model is that the escape time should vary in different subbands, since the mobility is inversely proportional to effective mass of charge carriers. In NTs, the effective mass $m^*$ of the second subband electrons and holes is twice that of first subband carriers ($m_2^* = 2m_1^*$) [17,22,23]. Due to the ratio of effective mass, the mobility $\mu$ (proportional to $1/m^*$) in the first subband should be twice that in the second subband $\varepsilon_2$. Including this with Equation (1) leads to an important experimental consequence: Carriers that are optically excited into the second subband (with effective mass $m_2^* = 2m_1^*$) of the PN junction should take longer than first subband carriers to accelerate out of the junction, assuming the scattering times are the same. Using ultrafast photocurrent measurements, we can probe the escape time of electrons and holes above and below $E_{22}$ and test this hypothesis.

Figure 4 shows measurements of the double pulse photocurrent vs. time delay in forward bias above and below $E_{22}$ for device 2. We observe that the normalized photocurrent above $E_{22}$ (blue data) decays with a time constant of $\tau_2 \sim 2.2$ ps, while the photocurrent below $E_{22}$ (red data) decays within $\tau_1 \sim 1.3$ ps. We plot the inverse decay times as a function of $V_{SD}$ for photon energies above (blue) and below (red) $E_{22}$. Similar to device 1 (Figure 3(c)), both data sets extrapolate to a $V_{SD}$ value consistent with the open circuit voltage $V_{OC} = 0.5$ V. However, while $1/\tau$ indeed scales linearly with $V_{SD}$, it exhibits a much steeper decent for $E_{PH} < E_{22}$. We fit both data sets and calculate the ratio of the extracted lifetimes and find $\tau_2 / \tau_1 \sim 1.7$, consistent with our hypothesis.

Finally, we consider high reverse bias region of Fig 3(c). The escape time becomes shorter and approaches a constant value $\tau_0$. To understand this behavior, we can compare the escape time $\tau$ to the average time between scattering events $\tau_S$. If the average time between scattering events $\tau_S = l / v_F$ is less than the escape time $\tau$, then carriers undergo diffusive transport through the junction. This is observed in forward bias. However, if $\tau_S \geq \tau$, then carriers may escape the junction without scattering and the escape time approaches the ballistic limit. In this limit, the transit time for a ballistic carrier across half of the PN junction ($L \sim 0.5$ μm) would exhibit crossover behavior to a constant escape time $\tau_0 = L / v_F \sim 0.5$ ps at sufficiently high reverse bias. This crossover behavior is indeed observed (solid blue line Figure 3(c)). However, the measured escape time is close to the experimental resolution of 0.4 ps, so further measurements with higher temporal resolution are needed to definitively confirm ballistic transport. Note that ballistic transport in reverse bias is consistent with previous findings in which $\varepsilon_2$ electrons and holes undergo highly efficient impact excitation resulting in multiple e-h pairs [2,24].

In summary, we have reported the first ultrafast photocurrent measurements that access the dynamics of electrons and holes in an individual nanotube PN junction. These experiments open the door to future photocurrent studies exploring aspects of NT optoelectronic response that have previously been probed only through optical measurements, including electron-hole recombination, phonon relaxation, and photoluminescence at various temperatures and photon energies. Our technique will open the door for more detailed measurements of multiple electron-hole pair generation [1-4] and electron-hole recombination [25,26] in other individual nanoscale devices that incorporate nanotubes, graphene, semiconductor nanowires and nanocrystal quantum dots.

**Figure Captions**

FIG. 1. Experimental apparatus and photocurrent characteristics of the NT PN junction. (a) Experimental apparatus: M1 translating mirror, M2 fixed mirror, BS beamsplitter. (b) $I$-$V_{SD}$ characteristics at $T$ = 40 K and $E_{PH}$ = 1.51 eV, for device 1 with open circuit voltage $V_{OC}$ = $E_{11}/e$ = 0.48 V. (c) Photocurrent vs. photon energy at $V_{SD}$ = 0.25 V. The top axis has been divided by $V_{OC}$ to assign the $E_{22}$ peak.

FIG. 2. Single pulse photocurrent of the NT PN junction. (a) Single pulse photocurrent vs. optical intensity at $T$ = 40 K, $E_{PH}$ = 1.51 eV and $V_{SD}$ = 0 V for device 1. Inset, single pulse photocurrent divided by the elementary charge $e$ and the repetition rate of the laser $f$ vs. optical intensity. (b) Same data as (a) in log-log scale.

FIG. 3. Double pulse photocurrent of the NT PN junction. (a) Photocurrent vs. time delay between two pulses at $V_{SD}$ = 0 V at increasing intensities ($n$ = 5, 11, and 26 x $10^{12}$ photons per pulse/cm$^2$ from top to bottom) for the same device and conditions as Figure 2. (b) Normalized photocurrent vs. time delay at $V_{SD}$ = 0 V (solid circles) and $V_{SD}$ = -0.3 V (open circles). (c) Extracted decay constant vs. $V_{SD}$. The red dashed line corresponds to the experimental resolution limit and the blue dashed line labels $\tau_0$ = 0.5 ps. Inset, Same data plotted as inverse decay constant $1/\tau$ vs. $V_{SD}$. The high reverse bias decay constant data is not shown in the inset. (d) Schematic of the escape time model for electrons and holes in the PN junction. Electrons (and holes, not shown) photo-excited at the center of the device travel a distance $L$ to escape the junction.

FIG. 4. Double pulse photocurrent of the NT PN junction as a function of photon energy. Inset, Normalized photocurrent vs. time delay at $E_{PH}$ = 1.51 eV (blue) and $E_{PH}$ = 0.85 eV (red) at $V_{SD}$ = 0.25 V for device 2: same device geometry and $V_{OC}$ = 0.5 V. Main panel: extracted inverse decay constants as a function of $V_{SD}$ for $E_{PH}$ = 1.51 eV (blue) and $E_{PH}$ = 0.85 eV (red) with linear fits to the data.

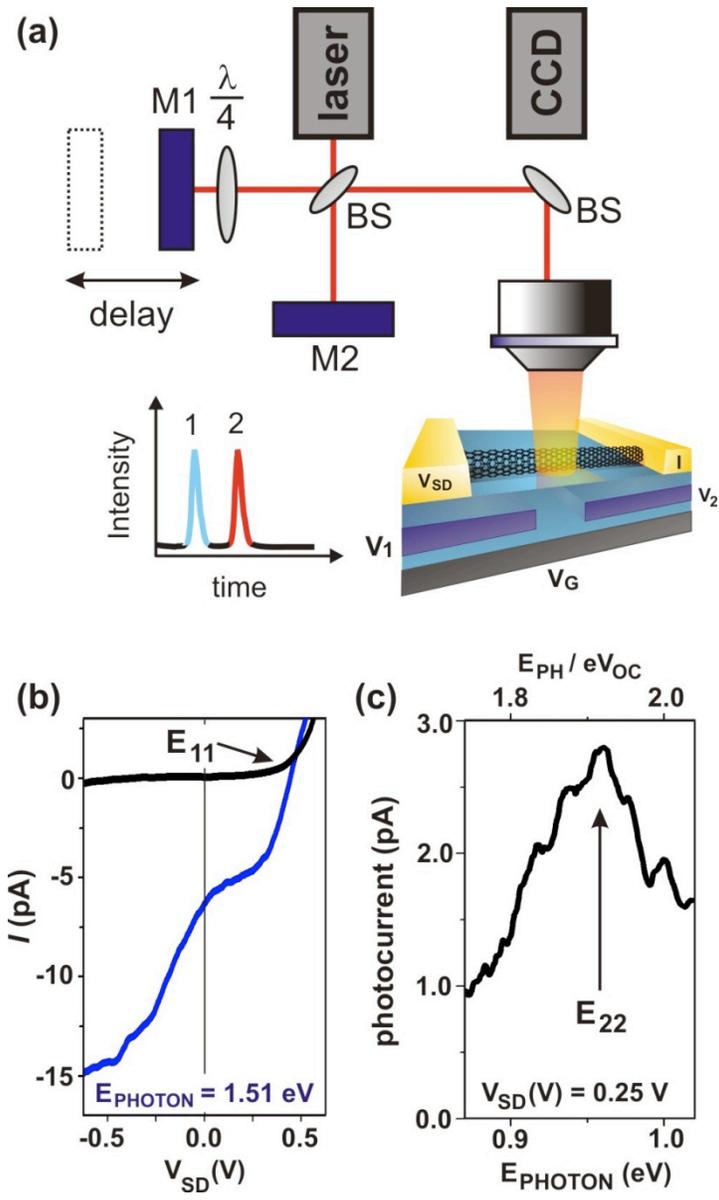

FIG. 2

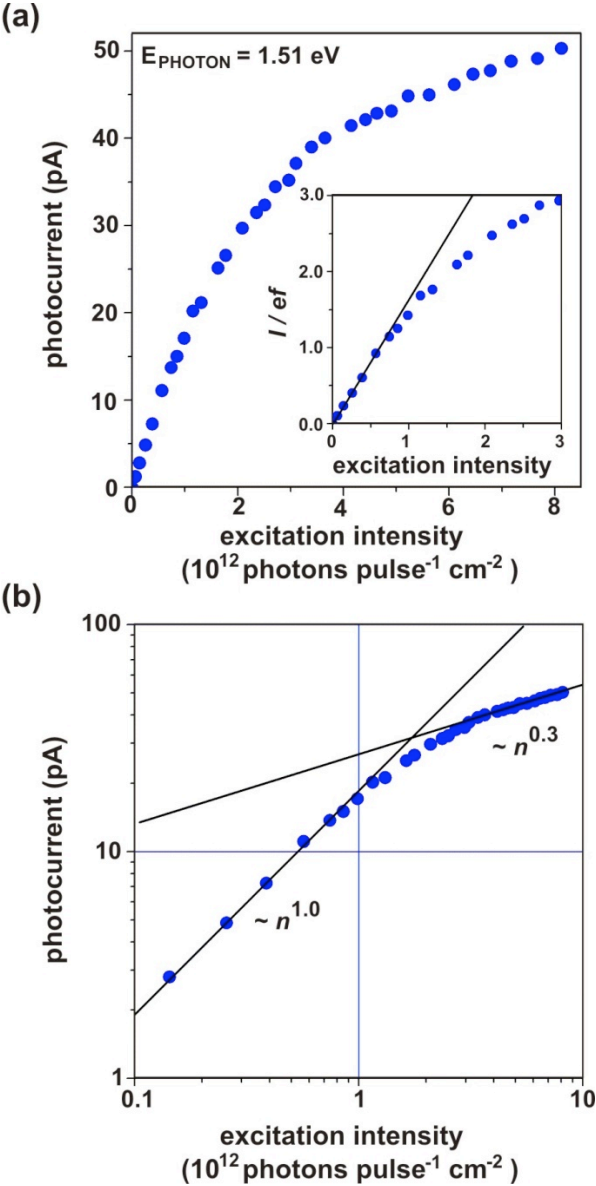

FIG. 3

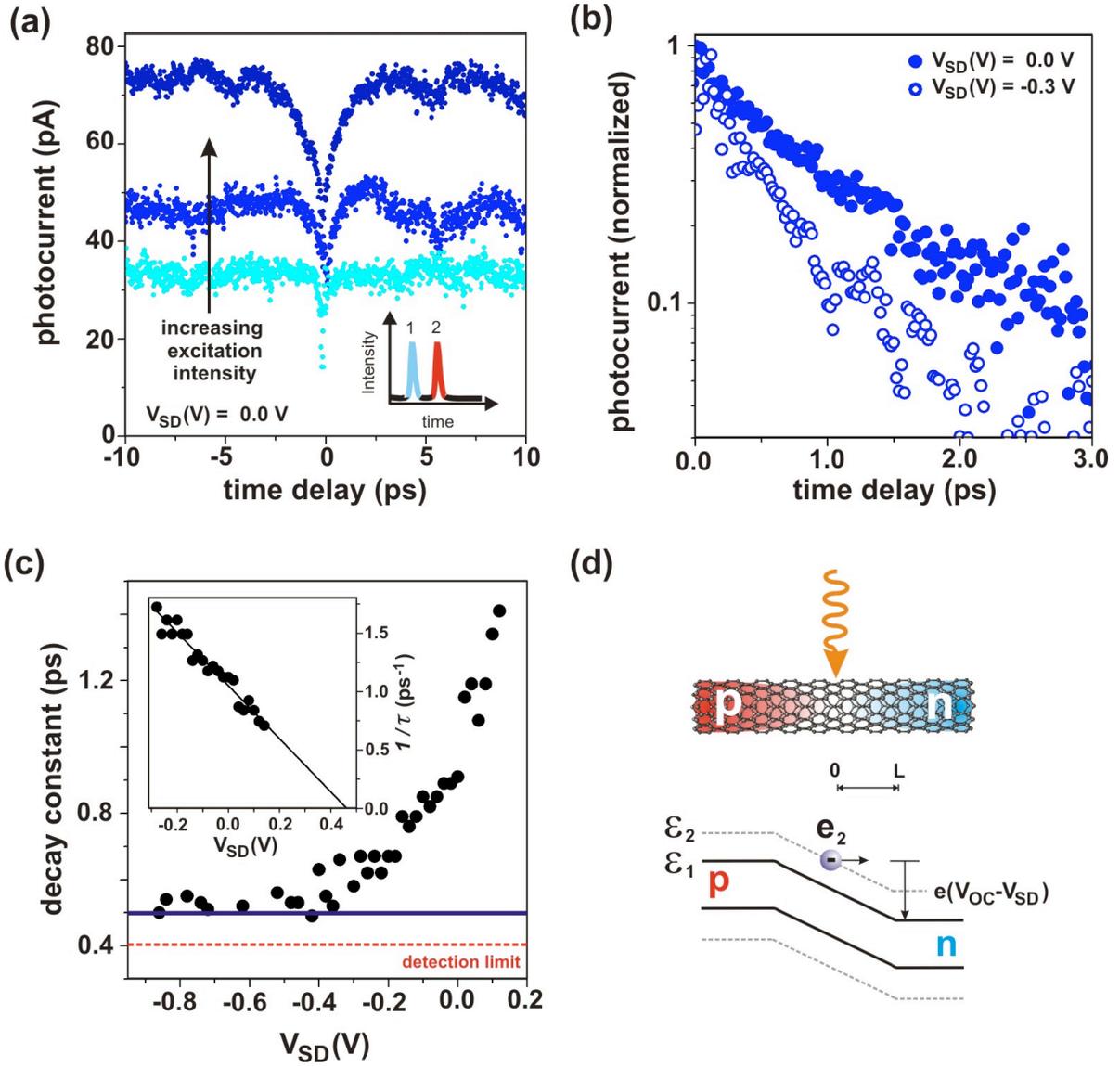

FIG. 4

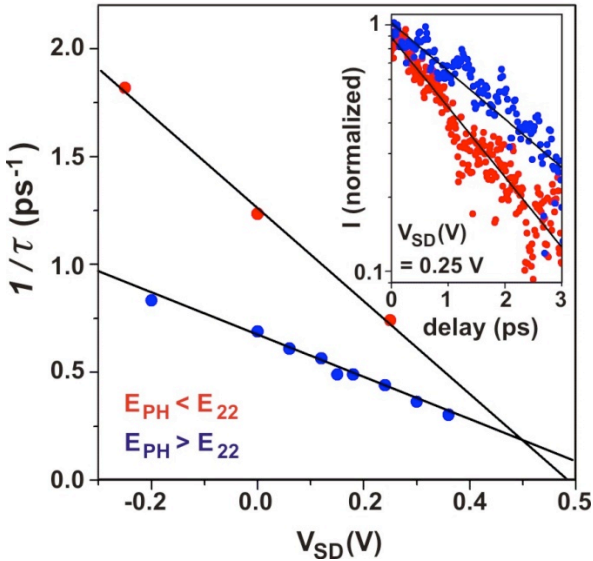